# A True AR Authoring Tool for Interactive Virtual Museums

Efstratios Geronikolakis[1,3], Paul Zikas[1,3], Steve Kateros[1,3], Nick Lydatakis[1,3], Stelios Georgiou[1,3], Mike Kentros[1,3], George Papagiannakis[1,2,3]

[1] Foundation for Research and Technology Hellas, 100 N. Plastira Street, 70013 Heraklion, Greece

[2] Computer Science Department, University of Crete, Voutes Campus, 70013 Heraklion, Greece

[3] ORamaVR, 100 N. Plastira Street, 70013 Heraklion, Greece
{stratos, paul, steve, nick, stelios, mike, george.papagiannakis}@oramavr.com

## Abstract

In this work, a new and innovative way of spatial computing that appeared recently in the bibliography called True Augmented Reality (AR), is employed in cultural heritage preservation. This innovation could be adapted by the Virtual Museums of the future to enhance the quality of experience. It emphasises, the fact that a visitor will not be able to tell, at a first glance, if the artefact that he/she is looking at is real or not and it is expected to draw the visitors' interest. True AR is not limited to artefacts but extends even to buildings or life-sized character simulations of statues. It provides the best visual quality possible so that the users will not be able to tell the real objects from the augmented ones. Such applications can be beneficial for future museums, as with True AR, 3D models of various exhibits, monuments, statues, characters and buildings can be reconstructed and presented to the visitors in a realistic and innovative way. We also propose our Virtual Reality Sample application, a True AR playground featuring basic components and tools for generating interactive Virtual Museum applications, alongside a 3D reconstructed character (the priest of Asinou church) facilitating the storyteller of the augmented experience.

**Keywords: Augmented Reality, Cultural Heritage, Virtual Museums, Gamification**

# Introduction

Augmented Reality has revolutionized many fields in the industry, from medical planning, to educational and training tools. It enhances the physical world with holographic assets, extending the possibilities of new learning and educational systems. Specifically, in the field of Digital Cultural Heritage, the rendering quality needs to be as high as possible to properly illuminate the artifacts, buildings or even characters, highlighting their natural beauty.

In this project, we followed a holistic approach generating True AR experiences by documenting both tangible and intangible heritage. To develop the holographic application, we exploited our M.A.G.E.S platform [11] as the core system architecture. This platform started as a tool to recreate psychomotor scenarios in VR for surgeons to master their skills but quickly expanded into other sectors like emergency scenarios, training courses and mechanical solutions. With this project, we aim to expand the capabilities of our system to support Cultural Heritage applications for educational purposes [15]. Our system generates realistic, gamified and interactive applications though Rapid Prototyping and Visual Scripting methodologies, forming an SDK for development of educational projects. By detecting the main restrictions that AR and HoloLens has, we took appropriate actions to transfer this work to AR, in order for the two versions to be as "close" as possible. The fact that someone is able to see objects or characters in AR and interact with them using their hands, increases the realism of AR (since a highly important element of realism is interaction as well, apart from the appearance of the characters, illumination etc.), thus getting us closer to our goal, which is True AR.

## 1. Previous work

### 1.1 The M.A.G.E.S platform

In this project, we exploited our award winning M.A.G.E.S. platform to generate a gamified AR scenario as an example of a holographic Virtual Museum [16]. This system can generate a fail-safe, realistic environment for surgeons to practice on VR scenarios, extending their skills in an affordable and portable solution. The key component of M.A.G.E.S platform lies on the customizable SDK platform able to generate educational training scenarios and immersive experiences with minimal adaptations and code-free due to the **Visual Scripting** and **Rapid Prototyping** mechanics.

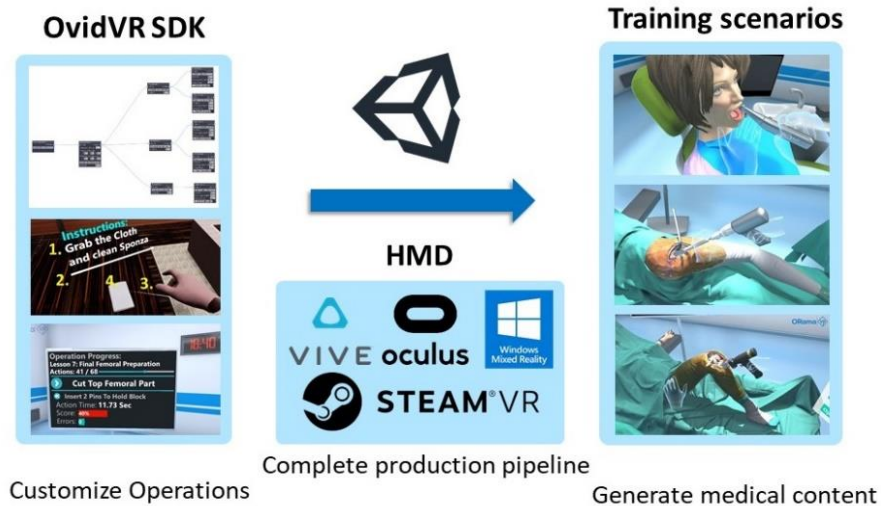

**Fig. 1** The M.A.G.E.S. flow diagram

The Visual Scripting component offers a node-based interface to generate VR behaviors and interactive tasks following a consistent educational pipeline. Utilizing this authoring tool, we developed a variety of medical training simulations, among them Total Knee and Hip Arthroplasties, a Dental Implant placement, an Endotracheal Intubation surgery and emergency medical scenarios (car-motorbike evacuation, trauma operations).

Rapid Prototyping of training scenarios reflects the classification methods to extract and break down complex educational pipelines into reusable and flexible building blocks. After documenting fundamental interaction tasks (insertion of objects/tools into highlighted areas, use of tools to complete an action, removal of objects using other tools etc.) we integrated those features into the M.A.G.E.S. platform forming a library of prototyped VR Actions in a user-friendly format, capable to generate more complex scenarios. Inspired from software design patters, we developed new VR design patterns, following interactive principles to replicate natural behaviors from real life into the VR world.

## 1.2 Holographic Virtual Museums

Since the advancement of holographic technology, AR headsets are evolving including interactive features like gesture and voice recognition, as well as improvements on resolution and FOV. In addition, untethered AR headsets paved the way for mobile experiences without the need of external processing power from a PC. Such embedded systems, facilitate great tools to represent virtual museums [8] due to their lack of cables and enhanced interactive capabilities. Virtual Museums are institutional centers in the service of society, open to the public for ac-quiring and exhibiting the tangible and intangible heritage of humanity for the purposes of education, study and enjoyment. In addition, True Augmented Reality has recently been defined to be a modification of the user's perception of their surroundings that cannot be detected by the user [13] due to their realism. Virtual characters and objects should blend with their surroundings, achieving the "suspension of disbelief".

In recent years, many approaches on holographic cultural heritage applications emerged, each one focusing on a different aspect of representing the holographic exhibits within the real environment. A published survey [5] investigated the impact of Virtual and Augmented reality on the overall visitor experience in museums, highlighting the social presence of AR environments. [9] presented a comparison of the latest methods for rapid reconstruction of real humans using as input RGB and RGB-D images. They also introduce a complete pipeline to produce highly realistic reconstructions of virtual characters and digital asserts suitable for VR and AR applications. Another project [2] discussed the development of a cross/augmented reality application for the Industrial Museum and Cultural Center in the region of Thessaloniki, Greece, to promote the preservation of CH in the area of Central Macedonia. The presented Mixed Reality application, integrates ARKit and ARCore to implement portal-based AR virtual museum along with a gamified tour guidance and exploration of the museum's interior. The storytelling factor of the application is dominant, focusing on the history of the museum, previous expeditions and its impact to the society of Thessaloniki in micro and macro level. Storytelling, Presence and Gamification are three very important fields that should be taken into account, when creating an MR application for cultural heritage. [10] presented a comparison of existing MR methods for virtual museums and pointed out the importance of these three fields for applications that contribute to the preservation of cultural heritage [3]. Moreover, in [4] fundamental elements for MR applications alongside examples are presented.

Another recent example [19] presented two Mixed Reality Serious Games in VR and AR comparing the two technologies over their capabilities and design principles. Both applications showcased the ancient palace of Knossos in Minoan Crete, Greece through interactive mini games and a virtual/holographic tour of the archeological site using Meta AR glasses. [1] successfully published an AR application for visualizing restored ancient artifacts based on algorithm that addresses geometric constraints of fragments to rebuild the object from the available parts.

### 1.3 Platforms for Gamified Content Creation

Authoring tools and other content creation platforms emerged in recent years to fulfill the need for interactive MR applications. BricklAyeR [14] is a collaborative platform designed for users with limited programming skills that allows the creation of Intelligent Environments through a building-block interface. ExProtoVAR [12] is a lightweight tool to create interactive virtual prototypes of AR applications designed for non-programmers lacking experience with AR interfaces. RadEd [17] features a new web-based teaching framework with an integrated smart editor to create case-based exercises for image interaction, such as taking measurements, attaching labels and select specific parts of the image. It facilitates a framework as an additional tool in complex training courses like radiology. ARTIST [6] is a platform which provides methods and tools for real-time interaction between human and non-human characters to generate reusable, low cost and optimized MR experiences. Its aim is to develop a code-free system for the deployment and implementation of MR content, while using semantically data from heterogeneous resources. The mentioned solutions provide developing environments to generate MR experiences, however they lack of advanced authoring tools and educational curriculum to support advanced educational - training scenarios.

## 2. Mixed Reality Sample app application

Our Sample app is a MR application, in which users are presented with basic examples of all the functionalities that our SDK supports. It is considered to be a playground for MR. We consider it as a room, where the users can experiment with the basic mechanics of our SDK, try them, and even create their own using our tools. In addition, these mechanics can be applied to other scenes as well, as they are not bound only on the specific application. The users can experience simple examples of different mechanics interacting with many objects in the scene (pick them up, hold them, even throw them), thanks to our "interactable item" utility that our SDK provides. They can use this functionality to possible objects that they will create, in order to interact with them and be able to move them around the scene with their virtual hands.

Another interesting mechanic that our SDK contains, is the virtual hands. They are automatically set up to follow and respond correctly to different kinds of controllers (Oculus, Vive, Mixed Reality, etc.). Users can easily set these hands to interact with objects of their choice. The hands are animated to perform the appropriate gesture that someone would perform in real life when the corresponding controller button is pressed. This functionality increases the realism and thus the feeling of presence in the scene.

An Inverse Kinematic (IK) object, a lamp, can be found in our MR scene, on the table. The users are free to interact with it, like any other movable objects in the scene and note how the IK mechanic works on the lamp (the base of the lamp moves with respect to the top of the lamp and vice versa), adding a great value of realism. In addition, users can use the presented IK mechanism to objects of their choice, including not only "lifeless" objects but virtual characters as well.

The Sample app application, offers a variety of interactive tasks (Actions). There is a variety of Actions where users can experiment with. Based on these Actions, users are able to create their own custom actions in the same or a new project simply by importing the SDK to their new project. Moreover, it provides the users with some basic tools (scalpel, mallet, scissors), which are in the scene during the whole operation and can be used to perform different kinds of actions. Based on these tools the users can even create their own tools to accomplish possible new actions that they will generate, according to their needs.

For the completion of some actions, (insertion of objects into specific positions) extra information is needed to inform user how to procced and execute this interactive task. This is done with another feature that our SDK provides, which is the *holograms*. They can be easily set up and can be used to indicate the correct position and rotation of an object during an "insert action", or the movement that the users have to do with an object in order to complete a "use action" successfully. The types of actions mentioned here are described later in this work.

A strong feature of our SDK is the UIs and the aidlines that it provides. Users can easily create their own UIs, based on our prototypes as well adding advanced functionalities to occur after pressing a UI button. The UIs are fully supported and interact with the virtual hands. The aidlines are mainly used to guide inexperienced users into tricky tasks. They include arrows, which can point to the preferred direction, followed by a message to inform the users of what they need to do. A typical example would be the aidline pointing to a tool and a message telling the users to pick up the specific tool and use it in a proper way.

```csharp
/// <summary>
/// Example of Insert Action
/// </summary>
public class AssembleSponzaPartOfAction : InsertAction
{
    /// <summary>
    /// Initialize method overrides base.Initialize() and sets the prefab user will insert
    /// </summary>
    public override void Initialize()
    {
        //Set Prefab to insert
        //First Argument: Interactable prefab
        //Second Argument: Final prefab
        //Third Argument: Hologram
        SetInsertPrefab("Lesson0/Stage1/Action0/SponzaInteractable", "Lesson0/Stage1/Action0/SponzaFinal");
        SetHoloObject("Lesson0/Stage1/Action0/Hologram/HologramSponzaFinal");
        SetAidLine("AidLine_Decision");

        base.Initialize();
    }
}
```

**Fig. 2** A basic example of an "insert action"

In Fig. 2 above, a basic example of the "Insert Action" for Sponza can be seen. The class inherits the InsertAction interface and implements the Initialize method, which is called at the beginning of the action. The method SetInsertPrefab, instantiates the interactable 3D model of Sponza in the scene and sets its final position to be the one indicated by the SponzaFinal prefab in this case. SetHoloObject generates a hologram (of the Sponza model, in this example) to the position where the interactable model should be set. Finally, the SetAidLine method generates the aidline prefab, the name of which is passed as an argument that becomes visible in the scene. Based on this prototype, the users can create their own actions from any 3D models, prefabs, holograms, aidlines etc.

Our UIs also offer another great feature. They can be used as notifications, providing helpful information to the users, as warnings, to inform them that they are possibly doing (or already did) something wrong (which may have not severe consequences but it is advisable to avoid it) and as errors to let them know that their actions caused a severe/critical error. With only a few lines of code, the users can generate whichever type of UI notification they prefer alongside a message.

Our Sample app application also supports a multiplayer collaborative environment. Many users (up to 7+) can join a virtual room, in order to collaborate with each other and complete actions together. To support a large number of participants in the virtual environment, we implemented our custom Conformal Geometric Algebra (CGA) GPU interpolation engine to reduce the data transfer on the network. Furthermore, thanks to dual quaternions, we accomplish smoother translations and rotations for our virtual objects. Since these functionalities are a part of our SDK, any user will be able to use them to create their online sessions of their application, without encountering many difficulties.

The application described in this work, is an AR version of this playground, the main objective of which is to help developers get started with the basic components of our SDK.

## 3. Integrate AR features into M.A.G.E.S. platform

The M.A.G.E.S. platform was designed for VR environments, thus integrating AR support as a part of the SDK was a challenging task. Virtual Reality reflects a fully digital environment to enhance immersion and embodiment, presenting a different reality through the virtual environment. On the other hand, Augmented Reality blends the virtual and the real world through the selective rendering of holographic assets. It is important to pay attention to the augmentations we add on top of the real environment, in order for them to blend well with their surroundings, obey the laws of gravity and match the environmental illumination. As it seems, AR and VR have not much in common. They present the digital content following different principles and design patterns. The figure below illustrates a screenshot from the VR application featuring the interaction with the North Gate of the palace of Knossos.

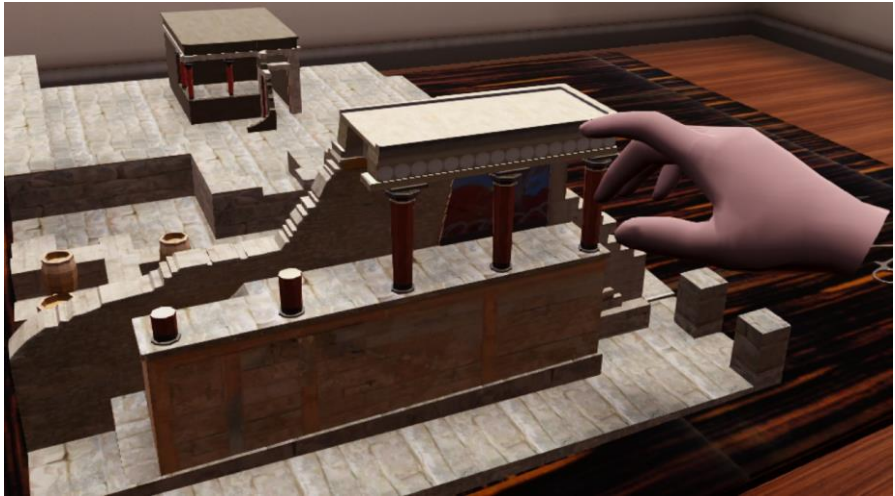

**Fig. 3** Interacting with the north gate of Knossos Palace

Our goal is to transform our platform into a cross-reality SDK capable to integrate multiple technologies (AR/VR/MR) within the same system. In the following sections we will discuss how we integrated AR support into a platform designed for VR.

### 3.1 The Modular Device Controller

The initial challenge is to design the system to manage and select the deployed device. We want our application to be cross-platform, able to generate executables for different devices and operating systems. M.A.G.E.S was built on Unity3D game engine to enable this feature. Unity may support exporting to different platforms but we have to change specific application-wise parameters and design principles to truly support this mechanic. Our goal was to develop a modular platform capable to support different realities without any parametrization from the developer's side, thus we had to integrate the AR support to the core engine of our system.

Our architecture contains different components to support multiple devices and technologies. We manage the deployed devices through the device controller module, which implements the callbacks for different buttons, analog sticks and gestures. As an example, we integrated SteamVR into our device controller module to support all the compatible VR headsets like Oculus, HTC VIVE, Microsoft Mixed Reality and many more. However, HoloLens does not support SteamVR, instead they offer HoloToolKit as the native API to manage the device. To integrate HoloLens into our platform, we implemented the HoloLens controller, a module which derives from the generic device controller and manages the callbacks from the HoloLens controller. However, HoloLens does not have a physical controller. Instead, users interact with the holographic environment using hand gestures. For this reason, we implemented HoloLens controller to support all the gestures and interactions from HoloToolKit. The diagram below illustrates our modular device controller architecture which supports input from various headsets and technologies.

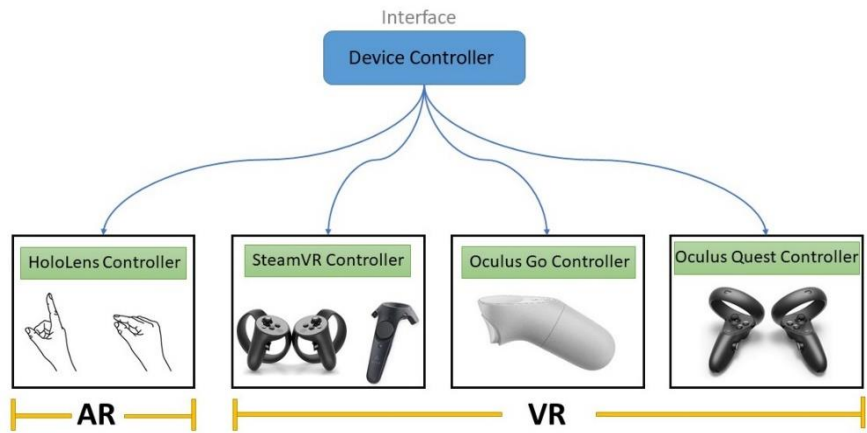

**Fig. 4** The Device Controller implementation diagram featuring the AR controller (left) and the VR controllers (right)

This implementation offers great flexibility over the platform enabling the project to run in different devices with minimal changes. To utilize the modular device controller, right before building the executable, we need to select the device in which we are deploying the application. Device controller is represented as a component in the Unity3D editor, thus we can select the deployed device from a convenient drop down menu we designed for this reason.

3.2 Interaction with holographic objects

The M.A.G.E.S. platform gives user the ability to interact with the 3D objects and the virtual environment through the interaction module we integrated. To support runtime interaction with the virtual objects, we integrated NewtonVR physics system as the main engine to handle the manipulation of 3D assets. The interaction physics engine of NewtonVR is velocity based, which means that interaction is not relied on parenting objects to user's hand but the objects are connected to the hands according to their current velocity. This approach gives the sensation of a more natural movement than the parenting mechanism. NewtonVR has a port for Unity engine with an active community and a significant number of successful VR applications.

Our interaction module is closely related to the device controller to support different headsets and devices. To integrate HoloLens platform into ours, we had to integrate the interaction system of HoloToolKit with NewtonVR. Cur-rently, the most common way to interact with holographic objects through Ho-loLens is by using the pinch gesture to grab a hologram and change its position within the virtual environment. This method utilizes a simple parenting mechanic to grab the object just by switching the parenting of the object to be the user's hand position. This mechanic is simple, but offers limited functionality and poor user experience. To improve the parenting grab mechanic and to unify the interaction mechanic in our platform, we integrated the HoloLens gesture grabbing system to the NewtonVR system, that we already support in our platform. The diagram below illustrates the interaction module which manages the input from HoloLens to send feedback to the platform.

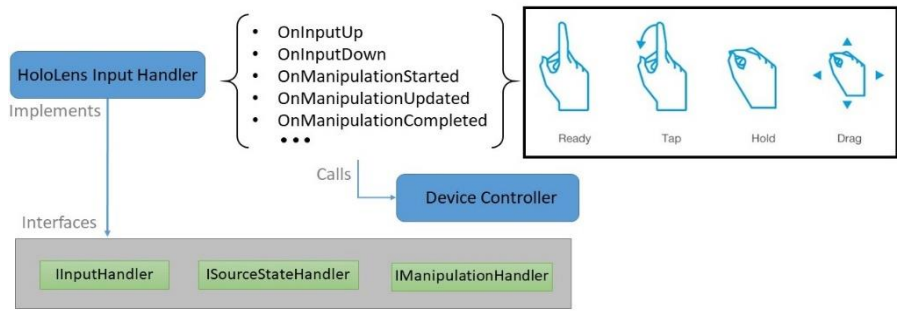

**Fig. 5** Diagram of the interaction module with additions to support HoloLens gestures

Our methodology was to create an intermediate module between the device controller and the HoloToolKit. This module is the HoloLens Input Handler. It implements three interfaces from HoloToolKit to link the inputs from Ho-loLens gestures and vocal controllers directly to our platform. For example, OnInputUp method is called when user is rising his pointer, indicating the first stage of tapping gesture. When HoloLens Input Handler recognizes this gesture, it automatically calls the appropriate Device Controller method to signal our platform that the gesture was performed.

### 3.3 Port an AR application to a VR system

At this point, we implemented the interaction module to handle the virtual assets with natural gestures. The next step is to reconstruct the augmented environment, where the user will interact with. The Sample App application was designed for a VR environment, thus the visualization of the 3D assets and the virtual room is fully digital. However, in AR applications, the rendered environment blends with the virtual and the real world since the augmentations do not cover the entire Field of View, but they are placed in key locations respecting physical objects.

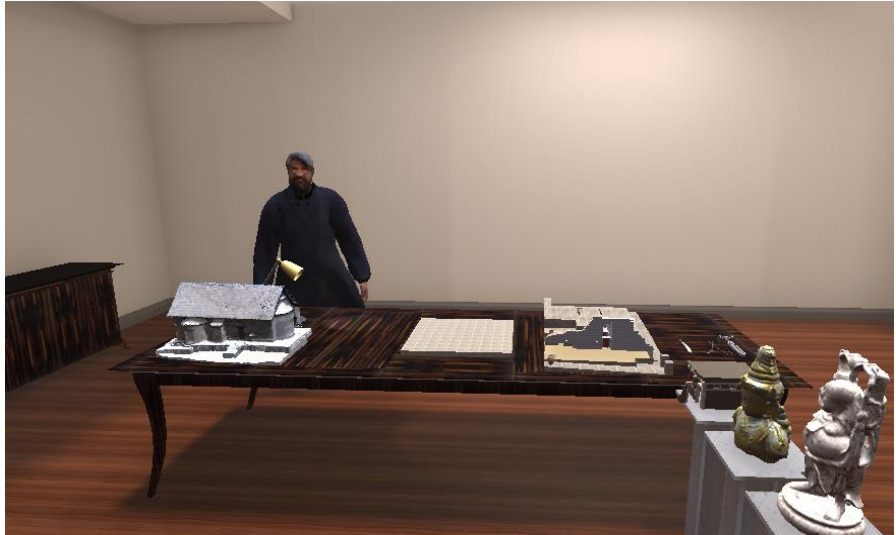

**Fig. 6** The VR version of our Sample Application

In more detail, to design the AR application, we have to keep only a small amount of digital assets and delete the majority of them to make room for the real environment. For this reason, we kept only the wooden table from the sample app room along with the interactable items and the priest of Asinou. In addition, to improve the realism of the holographic assets, we integrated a shadow plane under each object to replicate a real time shadow. This technique is simple enough but enhances the depth of field including an additional layer of illumination.

Another module we need to consider when changing the deployed medium (AR/VR) is the camera object, which represents the used HMD. For this reason, we integrated the HoloLens camera from the HoloToolKit into our system to support both cameras and technologies. In this way, developers can set the camera with a single click without the need to import any additional packages or libraries transforming sample app into a plug and play system.

## 4.   An Interactive AR Cultural Heritage Application

In this work, an innovative way of supporting AR in the VR version of our cultural heritage application is presented, in which the construction and the restoration of archaeological sites is simulated. This application combines education with entertainment featuring a serious game experience. The users, who complete this simulated procedure will learn about the buildings from within the archaeological sites, since they experienced their reconstruction/restoration. This is very important for the preservation of cultural heritage, because in case such an applica-tion is installed in the museum, the users will more likely want to visit the museum again, in order to try the application once more retaining more knowledge from their additional visit(s). For this reason, we enhanced the application with gamification elements [7] making the experience more appealing.

In this application, the construction of Knossos and the restoration of Sponza are simulated, but it can also be extended to support more monuments in the future. In terms of Knossos, users can pick up the indicated parts of Knossos (for which holographic representations are used to guide the users to the correct parts) and place them in the correct positions, again indicated by the use of holographic representations. In terms of Sponza, users have to restore the building of Sponza, by placing it in the annotated position and then executing the appropriate actions to restore the damaged building to perfect condition. If the user makes an error, it is indicated in the application by applying temporarily red color to the area of the error. All the actions that are referenced above, must be executed in a specific order, in order to complete the training scenario.

The featured types of actions in this application are described below:

- **Insert Action:** The user has to pick up an object and place it in the position indicated by the corresponding hologram.
- **Remove Action:** The user has to pick up the flashing object and move it away from its current position.
- **Tool Action:** The user has to execute a specific tool-driven action (cutting with scissors or scalpel, for instance).
- **Use Action:** The user has to execute an action using an object that is not a registered tool in our SDK (wiping over a surface with a cloth, for instance).

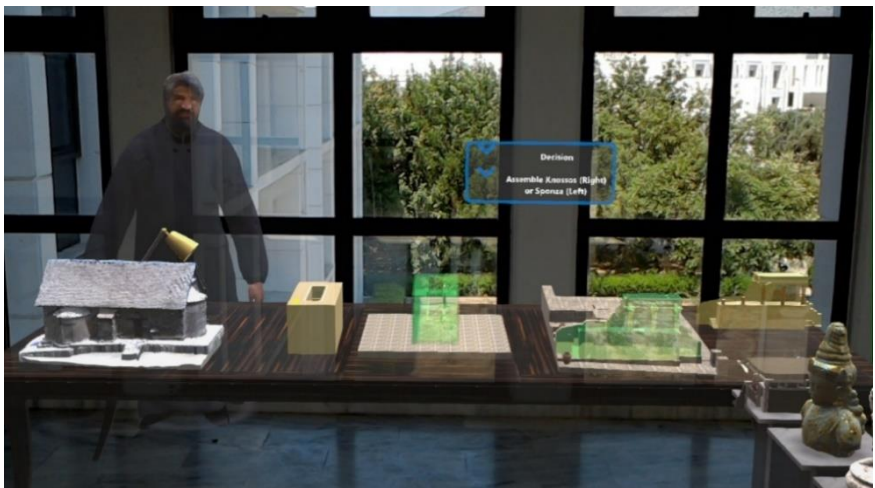

**Fig. 7** Main application. Restoration of Knossos (right) or Sponza (left)

Quizzes are also a part of the application. When starting the application, users are asked "Where is Sponza located?" and they have to choose one of the three proposed countries (each represented by its flag) by gazing at it and performing the tap gesture. Once the users choose an answer, both a visual (red/green color) as well as an audio feedback is given to them indicating whether they chose the correct answer or not.

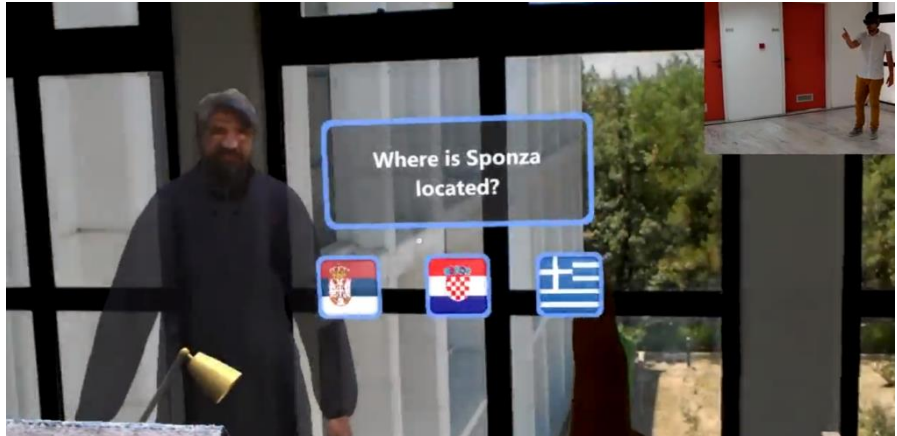

**Fig. 8** The user (top right) answering the question by performing the "tap" gesture

The application also features a life-sized priest, from the Asinou church, which is located in Cyprus. The priest stands behind the table all the time, watching the users executing the actions. Also, on the table in front of the priest, there is a miniature version of the Asinou church. The users can pick it up and inspect it by gazing at it and using the pinch gesture. Once the church is picked up, the priest notices it and starts speaking, informing users about the history of the church. As he speaks, the priest moves his arms and his body posture in general, to emphasize the important information in his sayings.

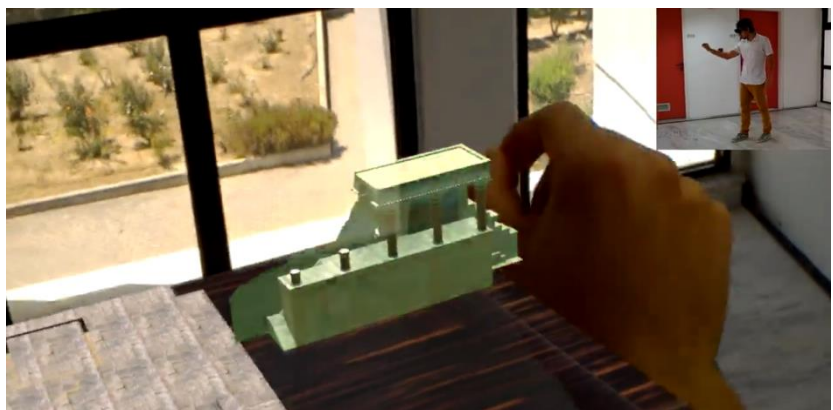

**Fig. 9** The user moving an object using the "pinch" gesture

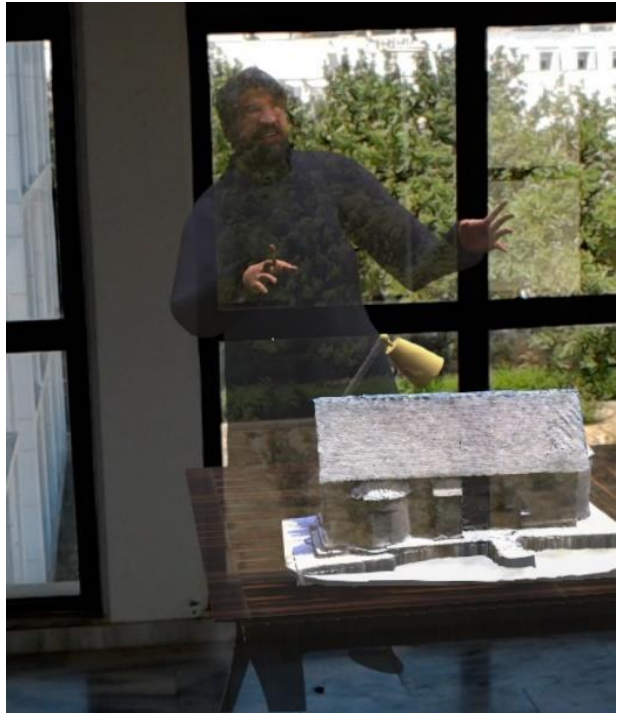

**Fig. 10** The priest telling history of the Asinou church (model on the table)

The priest mentioned above, is the real priest of the Asinou church. In fact, he was reconstructed by scanning the real priest with the occipital structure sensor, a sensor that connects to an iPad and is able to scan real 3D geometry [9]. After the scanning procedure, the model of the priest was improved with additional editing software (both in terms of geometry as well as texturing) to come correct scanning faults.

This application contributes a lot to the preservation of cultural heritage. Specifically, it provides the users with the chance to build or restore an archaeological monument themselves. By allowing the users to be actively involved to the reconstruction/restoration of the archaeological monuments, they gain even more knowledge by the end of the day as the application becomes more interesting and appealing.

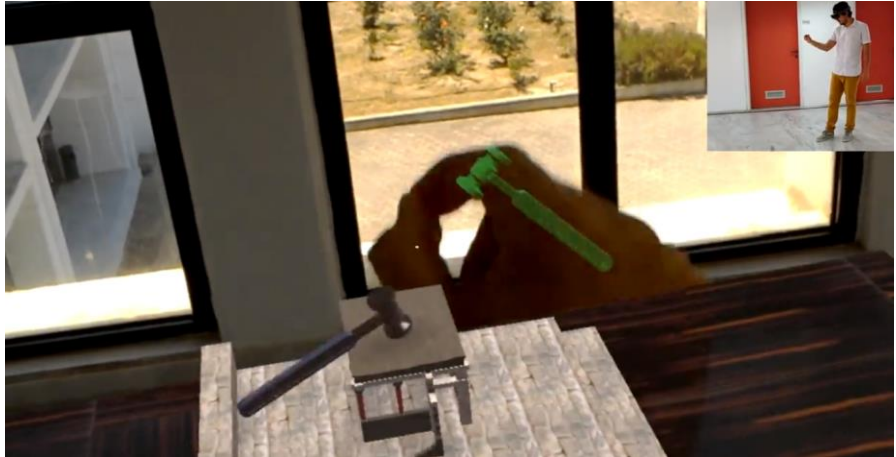

**Fig. 11** The user restoring the site by performing the action that the hologram indicates

The installation of gamified applications like this to museums, will lead to even greater attraction of people of younger age to museums. Nowadays, younger people believe that a visit to a museum is tiresome and only a few of them actually want to visit a museum. This is mostly because they do not have something, with which they will be able to interact. By offering interaction through immersive applications, the knowledge that museums offer will spread more quickly to younger ages, which will make it easier to preserve cultural heritage across generations.

## 5. Conclusions and Future Work

In this work, we presented a first integration of AR features in our SDK. By using the HoloToolKit for HoloLens and implementing an input handler for it that "communicated" with our Device Controller, we were able to use the gestures supported by HoloLens, in order to interact with different objects in our application. The main content of this application is the restoration of archaeological sites, in which the users can restore or reconstruct Sponza and Knossos respectively, by using gestures, which are mapped to the respective buttons of the controllers that were used in the VR mode. It is a great approach for the preservation of cultural heritage, since it provides the ability to be used in the real monuments, since the users are able to see the real world as well. By allowing, for instance, the visitors of the Knossos archaeological site to use this application on the site itself, during their visit, it will make their experience more interesting and fun, something that will increase the chance that they will visit Knossos again (in this specific example), or that they will recommend this site to others.

We also presented a 3D reconstructed virtual character for storytelling. This character tells the story of a monument (Asinou church in Cyprus), by using appropriate lips and body animations to make the interaction more realistic. This character, the priest, was reconstructed out of the real priest of the Asinou church. The fact that a virtual person exists in the application and can interact with the users, is an element that increases the feeling of presence and the realism of AR, bringing us one step closer to True AR.

During our work, we faced some limitations regarding the hardware of the holographic device. HoloLens has a rather small field of view, which does not allow the users to see the entire virtual scene, something not present in VR or in real life. Instead they are able to see the virtual scene through a small window (about 35 degrees), which repels them from fully immersing into the virtual experience. Another limitation is the processing power, as it is significantly lower than a desktop computer, in which the VR version of this application was running flawlessly. As a result, a scene that contains a large number of 3D objects (with complex geometry) may cause frame drops in HoloLens. Thus, it would be useful to create an algorithm/plugin in the future, which will be able to revise the complex geometry of a 3D object/model automatically, that it will be able to run in devices with less processing power. Lastly, we encountered a challenge during the port of actions that required tools, which respectively needed a specific button to be pressed in order for them to be used, i.e. scissors. A possible solution for this limitation is discussed later in this section.

There are many interesting and useful goals to be considered in the future. Since it is the first time that our application is transferred to AR, the walls, ceiling and floor that were surrounding the virtual room are removed, as they did not contribute in realizing the main concept and features of AR. For the future, it is worth considering a mechanism, which will be able to detect and do this automatically. By choosing, for instance, whether the project/scene is considered to be in AR or not, or by marking the objects of the scene that we would like to remain in the AR mode, the same version of the application will be able to run correctly both in AR and VR. Also, another feature that we would like to add, is a plugin, which will set up the scene automatically (setting up all the necessary objects for the reality that interests us at the moment). The programmer will only have to choose whether this is an AR project or not and the operating system that the project will run on, in order for it to set up all the required objects for the specific platform. This feature will be a meaningful addition to our SDK, as it will allow us to quickly build any project to whichever reality we want, without setting it "by hand".

Also, another important feature that we would like to add to the AR version is the online multiplayer part. In our SDK, more than one users can join the virtual room, in order to cooperate and reconstruct the building at the same time, by helping each other or with the guidance of an expert. That feature is missing from our AR version, and it is of high priority to support it in AR as well. Apart from the online cooperation of the users from anywhere in the globe, we would also like to try mixed online sessions with users that are using the VR version of the application and users using the AR version. The users in VR will be able to roam the room and execute the actions with AR users. This cross-reality online session is a great novelty both in the field of VR/AR, as well as cultural heritage preservation.

In the future we aim to conduct a qualitative evaluation survey to examine and document the capabilities of our system in real use. Our evaluation process will be based on [18] as this model reflects better our system functionalities and usage.

Finally, we aim to extend our SDK for HoloLens, in order to use voice commands, which are supported by HoloToolKit. Currently, the actions can be completed only by performing pinch and tap gestures. As HoloLens does not have any controllers, it is difficult to simulate the action where the users hold a tool with the grip button and press the trigger button to activate it at the same time. By supporting voice commands, the users will be able to hold a tool with their hand and say, for instance, a word like "use", in order to use the tool. The support of voice commands is not limited only to this, but can be more general, like giving the users the ability to execute the actions with their voice, or even more interestingly to start a conversation with the priest through dialogue-based interaction, asking him questions about the church, in order for him to provide the answers.

A video of this work is available here:
https://www.dropbox.com/s/eqzpsp0xbspzhik/TrueAR.mp4?dl=0